\documentclass{Interspeech}

\interspeechcameraready

\title{HiFiTTS-2: A Large-Scale High Bandwidth Speech Dataset}

\author[affiliation={1}]{Ryan}{Langman}
\author[affiliation={1}]{Xuesong}{Yang}
\author[affiliation={1}]{Paarth}{Neekhara}
\author[affiliation={1}]{Shehzeen}{Hussain}
\author[affiliation={1}]{Edresson}{Casanova}
\author[affiliation={1}]{Evelina}{Bakhturina}
\author[affiliation={1}]{Jason}{Li}

\affiliation{}{NVIDIA}{USA}
\email{\{rlangman, xueyang, pneekhara, shehzeenh, ecasanova, ebakhturina, jasoli\}@nvidia.com}

\usepackage{comment}
\usepackage{hyperref}
\usepackage[table]{xcolor}
\usepackage{colortbl}
\usepackage{subcaption}

\begin{document}

\maketitle

\begin{abstract}
    
    This paper introduces HiFiTTS-2, a large-scale speech dataset designed for high-bandwidth speech synthesis. The dataset is derived from LibriVox audiobooks, and contains approximately 36.7k hours of English speech for 22.05 kHz training, and 31.7k hours for 44.1 kHz training. We present our data processing pipeline, including bandwidth estimation, segmentation, text preprocessing, and multi-speaker detection. The dataset is accompanied by detailed utterance and audiobook metadata generated by our pipeline, enabling researchers to apply data quality filters to adapt the dataset to various use cases. Experimental results demonstrate that our data pipeline and resulting dataset can facilitate the training of high-quality, zero-shot text-to-speech (TTS) models at high bandwidths.
    
\end{abstract}

\section{Introduction}

Text-to-speech (TTS) and speech synthesis technology have advanced rapidly in recent years. Initially, TTS research focused on creating a small number of high-quality voices for applications like voice assistants. Datasets developed for this purpose, such as LJSpeech~\cite{ljspeech17} and HiFiTTS~\cite{bakhturina2021hifi}, typically contained large amounts of audio from one or a few speakers, emphasizing a need for clean or studio-quality audio at high frequency resolutions like 22.05 kHz or 44.1 kHz.

More recently, research has shifted towards zero-shot TTS, which aims to replicate the voice of a speaker not seen during training, using only a few seconds of reference audio. A common approach for zero-shot TTS is training an autoregressive language model to predict audio tokens produced by a neural audio codec~\cite{chen2025neural,zhang2024speechtokenizer,neekhara2024improving,hussain2025koelttsenhancingllmbased,chen2024vall,casanova2024low,Peng2024VoiceCraftZS}. These approaches rely on large amounts of transcribed training data, which is challenging to obtain in the speech domain. Consequently, most recent research uses large volumes of low-quality 16 kHz audio~\cite{zhang2024speechtokenizer,Peng2024VoiceCraftZS,chen2024vall}. While these models achieve high performance for zero-shot TTS, they synthesize speech with bandwidth and subsequent quality too low for many real-world applications.

Modern TTS models have different data requirements compared to earlier models. While previous models often required clean or studio-quality audio~\cite{shen2018natural,ren2019fastspeech} due to their inability to accurately model noise or acoustics, current models can learn these speech characteristics through in-context learning~\cite{chen2025neural}. For this reason, our data processing does not focus on signal-to-noise ratio (SNR) metrics, which were emphasized in earlier works and datasets like HiFiTTS.

% There are a few common problems with using speech data obtained in the wild. 
Several challenges arise when using speech data from diverse real-world sources. There are significant legal complexities surrounding data use for applications like speech synthesis and voice-cloning. For instance, Emilia~\cite{he2024emilia}, a large-scale speech dataset, uses a non-commercial license due to its underlying sources having varied licenses and copyright. In contrast, our dataset is suitable for commercial use, as it derives from LibriVox~\cite{librivox} audiobooks in the public domain. We remove data from LibriVox speakers who have explicitly requested their recordings not be used for machine learning applications.

Another challenge comes from the common practice of storing audio in a standardized format regardless of how it was recorded. This often results in audio being upsampled to a higher resolution than its original recording. Many publicly available datasets with high sampling rates, such as CommonVoice~\cite{Pratap_2020}, Emilia~\cite{he2024emilia}, and DNS~\cite{dubey2024icassp}, contain such mixed-bandwidth audio. LibriVox audiobooks are stored at 24 kHz and 48 kHz, but datasets derived from them are often downsampled to 16 kHz, as seen in Librispeech~\cite{panayotov2015librispeech}, MLS~\cite{Pratap_2020}, and LibriLight~\cite{kahn2020libri}.
Mixed-bandwidth audio can be problematic for training speech models. For example, studies have shown that training audio codecs on mixed-bandwidth data results in attenuation of all audio information above a certain frequency threshold~\cite{kumar2024high,puvvada2024discrete}. Training with mixed-bandwidth data benefits from bandwidth labels, which can be used to filter data~\cite{kumar2024high}, be provided as input to models~\cite{mantena19bandwidth}, or used for bandwidth extension~\cite{su2021bandwidth}.

In this paper, we present HiFiTTS-2, a large-scale audiobook dataset designed for speech synthesis. Our contributions are as follows:
\begin{itemize}
    \item We present a data processing pipeline consisting of bandwidth estimation, segmentation, text processing and validation, and multi-speaker detection.
    \item We create and release a dataset\footnote{https://huggingface.co/datasets/nvidia/hifitts-2} with two subsets by applying this pipeline to LibriVox audiobooks: a 22.05 kHz subset with 36.7k hours of speech from 5k speakers, and a 44.1 kHz subset with 31.7k hours of speech from 4.6k speakers.
    \item We provide precomputed metadata at the utterance and audiobook chapter level, which users can use to apply quality control filters optimized for their individual use cases.
    \item We demonstrate the effectiveness of our data pipeline by conducting experiments with a state-of-the-art TTS model.
\end{itemize}

\section{Data Processing Pipeline}

\setlength{\tabcolsep}{4pt}
\begin{table*}[ht]
  \caption{Text preprocessing examples applied to original MLS input. The process restores punctuation and capitalization, normalizes text, and replaces formatting elements such as `nbsp' and `p p'.}
  \vspace{-2mm}
  \footnotesize
  \label{table:text_preprocess}
  \centering
  \begin{tabular}{>{\raggedright\arraybackslash}p{8cm} | >{\raggedright\arraybackslash}p{8cm}}
    \toprule
    Input Text & Preprocessed Text \\
    \midrule
    \colorbox{yellow!30}{beautifully-shaped} and coloured glass and saltcellars tankards \colorbox{yellow!30}{c} of gold and silver & \colorbox{green!20}{beautifully shaped} and coloured glass\colorbox{green!20}{,} and saltcellars\colorbox{green!20}{,} tankards\colorbox{green!20}{, et cetera} of gold and silver\colorbox{green!20}{.} \\
    \midrule
    at that moment of supreme anxiety\colorbox{yellow!30}{ nbsp it} is my purpose\colorbox{yellow!30}{ nbsp mr allen} read & at that moment of supreme anxiety\colorbox{green!20}{. `` `It} is my purpose\colorbox{green!20}{,' " mister Allen} read \\
    \midrule
    in his calculations\colorbox{yellow!30}{ p p rather} would \colorbox{yellow!30}{i} believe that \colorbox{yellow!30}{i} have been mistaken in the affection which \colorbox{yellow!30}{i} feel for him said \colorbox{yellow!30}{mrs evangelina} & in his calculations\colorbox{green!20}{. ``Rather} would \colorbox{green!20}{I} believe that \colorbox{green!20}{I} have been mistaken in the affection which \colorbox{green!20}{I} feel for him\colorbox{green!20}{,"} said \colorbox{green!20}{misses Evangelina} \\
    \bottomrule
\end{tabular}
\vspace{-4mm}
\end{table*}

We use the English subset of MLS~\cite{Pratap_2020} as the input to our pipeline. MLS is a multi-lingual dataset derived form LibriVox audiobooks, primarily designed for ASR model training. The English subset of MLS contains approximately 44.7k hours of speech from 5,574 speakers. However, several significant challenges arise when using the original dataset for training TTS models:
\begin{itemize}
    \item Text transcriptions lack punctuation and capitalization, which plays crucial roles in conveying prosodic features in TTS.
    \item All audio data is stored at a 16 kHz sampling rate, regardless of their original bandwidth.
    \item There are no utterances shorter than 10 seconds, with all utterances ranging from 10 to 20 seconds in length.
    \item Text transcriptions are taken directly from the audiobook text, which may differ slightly from the speaker's narration.
    \item Speaker labels may be unreliable, as some audiobooks are narrated by multiple speakers.
\end{itemize}
In the following sections, we describe our processing pipeline to address these issues. Processing steps are carried out in the order listed.

\subsection{Text Preprocessing}

We start with the utterances and their transcripts from the English subset of MLS. To recover punctuation and capitalization (PC) for the transcripts, we take two approaches:
\begin{enumerate}
    \item We download the original audiobook text for each chapter. After removing PC from the audiobook text, we check whether each MLS transcript matches a substring in the text. If a match is found, we replace the original MLS transcript with the corresponding audiobook text. This method successfully recovers PC for approximately 87\% of the original MLS transcripts.
    \item For the remaining 13\% of transcripts, we predict PC using NeMo DistilBERT~\cite{nemo_en_distilbert}.
\end{enumerate}
%First, we download the original audiobook text for each audiobook chapter. We remove PC from the audiobook text, then take each MLS transcript and check whether it matches a substring in the text. If a match is found, then we replace the original MLS transcript with the corresponding audiobook text. Doing this we successfully recover the PC for approximately 87\% of the original MLS transcripts. For the remaining 13\% of transcripts, we predict PC using NeMo DistilBERT~\cite{nemo_en_distilbert}. In our dataset we provide metadata specifying which utterance transcriptions are from the audiobook and which are from MLS with predicted PC.
In our dataset we provide metadata specifying which utterance transcriptions are from the audiobook and which are from MLS with predicted PC.
After restoring PC to the text, we attempt to remove various formatting information common in the downloaded audiobook text, such as HTML tags. Lastly, we normalize the text using NeMo text normalization~\cite{zhang2021nemo}. Table~\ref{table:text_preprocess} shows a few examples of text preprocessing.

\subsection{Audio Processing}

To process the audio for the dataset, we first download all of the original 48 kHz audiobook files using the LibriVox API. We downsample the audio to 44.1 kHz and convert files from MP3 to FLAC format. We apply energy-based silence trimming with a threshold of 50 dB, leaving at most 0.5 seconds of silence at the start and end of each utterance.

\subsection{Bandwidth Estimation}

We apply the bandwidth estimation approach from the HiFiTTS~\cite{bakhturina2021hifi} to the first 30 seconds of each audiobook. The bandwidth $f_{\text{max}}$ is estimated by using the mean of the power spectrum to find the highest frequency that has at least -50 dB level relative to the peak value of the spectrum, namely,
\begin{equation*}
    f_{\text{max}} = \max\left\{f \in [0, f_{\text{Nyquist}}] \, \bigg|\, 10 \log_{10} \left(\frac{P(f)}{P_{\text{peak}}}\right) \geq -50\, \text{dB}\right\}
\end{equation*}
where $P(f)$ is the power spectral density and $P_{\text{peak}}$ the maximum spectral power.
Utterances will have an estimated bandwidth that is at most its Nyquist frequency. So audio recorded at 16 kHz sampling rate will have estimated bandwidth less than or equal to 8 kHz and audio recorded at 24 kHz will have estimated bandwidth less than or equal to 12 kHz.

For the 22.05 kHz subset of our dataset, we filter for utterances with an estimated bandwidth of at least 11 kHz. This produces approximately 36.7k hours of full-bandwidth 22.05 kHz speech.
% We obtain audio suitable for 22.05 kHz training by filtering for utterances which have an estimated bandwidth of at least 11 kHz. This produces approximately 36.7k hours of full-bandwidth 22.05 kHz speech.

Processing 44.1 kHz audio introduces additional complexity. Most speech recorded at a sampling rate of 24 kHz or lower is close to full-bandwidth. However, speech recorded at higher sampling rates like 48 kHz often has little spectral information in the highest frequency bands, resulting in an estimated bandwidth that is significantly lower than its Nyquist frequency. This results in most data for 44.1 kHz or 48 kHz training being effectively mixed-bandwidth. Training models on mixed-bandwidth data is challenging, and research on it is relatively uncommon due to most large public datasets being 16 kHz sampling rate.

\begin{figure}[!ht]
    \vspace{-2mm}
    \centering
    \noindent{\includegraphics[width=0.7\columnwidth]{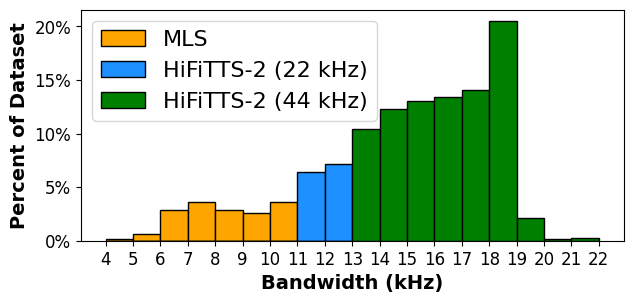}}
    \vspace{-2mm}
    \caption{Distribution of estimated bandwidth across utterances. Orange shows low-bandwidth utterances in MLS that are excluded from HiFiTTS-2. Green denotes high-bandwidth utterances present in all datasets. Blue denotes utterances included in the 22 kHz subset but absent from the 44 kHz subset.}
    \label{fig:bandwidth}
\vspace{-4mm}
\end{figure}

To address this, we filter 44.1 kHz audio using a bandwidth filter of 13 kHz, similar to HiFiTTS~\cite{bakhturina2021hifi}. This removes all audio recorded at a sampling rate of 24 kHz or lower. We assume that recording at sampling rates between 24 kHz and 44.1 kHz is very rare, even for data in-the-wild. This process yields approximately 31.7k hours of 44.1 kHz speech with bandwidth ranging from approximately 13 kHz to 22 kHz. We release this subset of the data, along with metadata containing the estimated bandwidth, to assist with future research on high bandwidth and mixed bandwidth modeling. The resulting bandwidth distribution is shown in Figure~\ref{fig:bandwidth}.

\subsection{Segmentation}

\begin{figure}[!ht]
    \vspace{-2mm}
    \centering
    \noindent{\includegraphics[width=0.7\columnwidth]{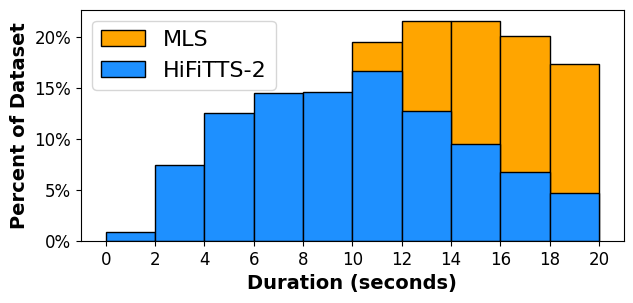}}
    % \caption{Distribution of utterance durations. Orange represents the distribution of durations in MLS, all between 10 and 20 seconds. Blue shows the distribution in HiFiTTS-2 after segmentation.}
    \vspace{-2mm}
    \caption{Distribution of utterance durations. Orange denotes the original distribution before segmentation.}
    \label{fig:duration}
\vspace{-4mm}
\end{figure}

Utterances in the MLS data range from 10 to 20 seconds in length. However, for TTS training a more balanced distribution of durations is desirable to prevent biases in model performance. To achieve this, we employ the NeMo Forced Aligner~\cite{rastorgueva2023nemo} with Parakeet-TDT-CTC~\cite{xu2023efficient} to generate alignment information. Using these alignments, we identify utterances containing a period followed by a pause of at least 0.08 seconds, ignoring periods within abbreviations.
We then split such utterances at the midpoint of the pause, similar to the segmentation methodology from  MLS~\cite{Pratap_2020}. If multiple qualifying pauses exist within an utterance, we select the longest one for splitting. In cases where multiple pauses have the same duration, we choose one at random. This process results in a duration distribution resembling a bell curve, as shown in Figure~\ref{fig:duration}.

\subsection{Text Validation}

\begin{figure}[!ht]
    \vspace{-2mm}
    \centering
    \begin{subfigure}[b]{1.0\columnwidth}\centering\noindent{\includegraphics[width=0.7\columnwidth]{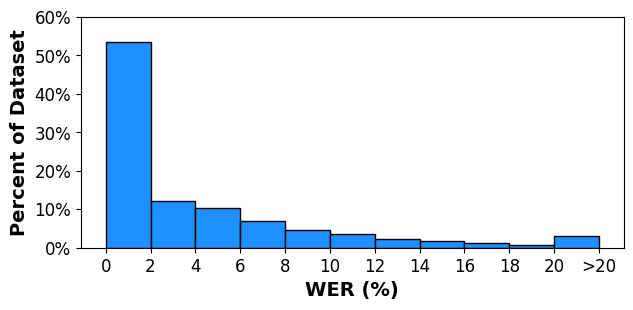}}
    \end{subfigure}
    \begin{subfigure}[b]{1.0\columnwidth}\centering\noindent{\includegraphics[width=0.7\columnwidth]{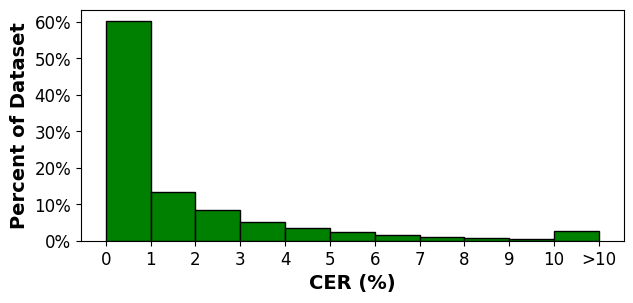}}
    \end{subfigure}
    \vspace{-4mm}
    \caption{Distribution of WER and CER in {HiFiTTS-2}.}
    \label{fig:wer_cer}
\vspace{-4mm}
\end{figure}

We validate the correctness of the final text using ASR. For all segmented utterances, we compute WER (word error rate) and CER (character error rate) using predicted transcriptions from Parakeet-TDT-CTC~\cite{xu2023efficient}. We filter out utterances with a CER of 100\% or greater, which correspond to utterances with incorrect transcripts or utterances containing only silence. Applying any higher threshold creates a trade-off between the volume and intelligibility of the training data. We provide the WER and CER as metadata in the dataset, allowing users to filter using a threshold appropriate for their use case. The utterance distribution of WER and CER is shown in Figure~\ref{fig:wer_cer}.

\subsection{Speaker Counts}
Some LibriVox audiobooks are narrated by multiple speakers, leading to utterances that are not always ideal for modeling purposes.
In order to identify and quantify the number of speakers present in each utterance, we employ Softformer~\cite{park2024sortformer}, a recent end-to-end neural model for speaker diarization. This model efficiently detects instances of multi-speaker speech, allowing us to filter and annotate such segments accordingly. Through this process, we identify approximately 104.4 hours of speech in the dataset as containing multiple speakers. We tag the number of speakers for each utterance as metadata in our dataset.

\section{Dataset Statistics}

HiFiTTS-2 comprises utterances from LibriVox audiobooks, with each utterance capped at 20 seconds in length. The dataset contains two subsets: 
\begin{enumerate}
    \item A full-bandwidth 22.05 kHz subset containing 36.7k hours of speech with 13.1M utterances from 5,013 speakers; 
    \item A 44.1 kHz subset with bandwidths above 13 kHz containing 31.7k hours of speech with 11.3M utterances from 4,631 speakers.
\end{enumerate}
These subsets are overlapping, with the 44.1 kHz being a strict subset of the 22.05 kHz.

Table~\ref{table:stats} provides a comparison between HiFiTTS-2 and other datasets commonly used for speech synthesis, highlighting its significant scale advantage.

We provide two sets of dev and test splits: one for speakers seen in the training data and the other one for unseen speakers not present in the training data. To ensure consistency and maximize data quality, we use the same utterances for both the 22.05 kHz and 44.1 kHz subsets, selecting only utterances with at least 13 kHz bandwidth, zero WERs, and a single speaker.

For unseen speakers, we use utterances produced by passing the MLS dev and test partitions through our data processing pipeline. This creates an unseen speaker dev set with 1098 utterances from 29 speakers and a test set with 968 utterances from 27 speakers.

For seen speakers, we create dev and test partitions each consisting of 1,000 utterances. These utterances were selected by sampling 50 speakers with 15 to 60 minutes of training audio, and sampling 20 utterances from each speaker such that utterances are uniformly distributed across gender, duration, and bandwidth.

%For seen speakers, we find speakers with 10 to 60 minutes of speech. This guarantees that seen speakers have sufficient training data, but not so much as to be impractical. From this set, we select uniformly at random 100 speakers, and 10 utterances from each speaker. This creates seen speaker dev and test sets, each containing 1,000 utterances from 100 speakers.

These carefully curated splits facilitate the robust evaluation of model performance, important for assessing the generalization of speech synthesis systems.
\setlength{\tabcolsep}{4pt}
\begin{table}[ht]
  \caption{Statistics for HiFiTTS-2 compared to other datasets.}
  \footnotesize
  \vspace{-2mm}
  \label{table:stats}
  \centering
  \begin{tabular}{l | r r r}
    \toprule
    Dataset & Sampling Rate & Hours & Speakers \\
    \midrule
    VCTK & 48 kHz & 44 & 110 \\
    HiFiTTS & 44.1 kHz & 292 & 10 \\
    LibriTTS & 24 kHz & 586 & 2,456 \\
    MLS (English) & 16 kHz & 44.7k & 5,574 \\
    HiFiTTS-2 (22 kHz) & 22.05 kHz & 36.7k & 5,013 \\
    HiFiTTS-2 (44 kHz) & 44.1 kHz & 31.7k & 4,631 \\
\bottomrule
\end{tabular}
\vspace{-4mm}
\end{table}

\begin{table*}[htbp]
    \centering
    % \caption{Koel-TTS results on seen and unseen speakers. Lower CER and WER indicate better intelligibility. Higher SSIM indicates better speaker similarity to the context audio. Higher SQUIM-MOS indicates better audio quality. The model trained on only HiFiTTS has no seen speakers in the test set.}
    % \caption{Koel-TTS performance on seen and unseen speakers. Lower CER and WER indicate better intelligibility. Higher SSIM indicates better speaker similarity to the context audio. Higher SQUIM-MOS indicates better audio quality. Note that the model trained exclusively on HiFiTTS has no seen speakers in the test set.}
    \caption{Koel-TTS results. Note that the model trained exclusively on HiFiTTS has no seen speakers in the test set.}
    \vspace{-2mm}
    \resizebox{\textwidth}{!}{
    \begin{tabular}{lrr|cccc|cccc}
    \toprule
    \multicolumn{3}{c}{} & \multicolumn{4}{c}{\emph{Seen Speakers}} & \multicolumn{4}{c}{\emph{Unseen Speakers}} \\
    \toprule
    Train Datasets & Dur~(hrs) & \#Spks & CER~(\%)$\downarrow$ & WER~(\%) $\downarrow$ & SSIM $\uparrow$ & SQUIM-MOS $\uparrow$ & CER(\%)$\downarrow$ & WER (\%) $\downarrow$ & SSIM $\uparrow$ & SQUIM-MOS $\uparrow$\\  
    \midrule
    \multicolumn{3}{c|}{\emph{Ground Truth (oracle)}}  & \emph{0.51}$\pm$\emph{0.00} & \emph{1.42}$\pm$\emph{0.00} & \emph{0.763}$\pm$\emph{0.000}      & \emph{4.616}$\pm$\emph{0.030} & \emph{0.80}$\pm$\emph{0.00} & \emph{1.83}$\pm$\emph{0.00} & \emph{0.771}$\pm$\emph{0.000} & \emph{4.588}$\pm$\emph{0.020}\\
    \midrule
    HiFiTTS & 283  & 10                & --   & --   & --    & --    & 6.97$\pm$0.37 & 10.38$\pm$0.36 & 0.059$\pm$0.004 & 4.380$\pm$0.005 \\
    %LibriTTS(2e-4) & 539  &   2,259               & 1.58$\pm$0.28 & 2.71$\pm$0.34 & 0.680$\pm$0.002 & 4.387$\pm$0.005 & 1.37$\pm$0.13 & 2.44$\pm$0.19 & 0.504$\pm$0.002 & 4.386$\pm$ 0.004 \\
    LibriTTS & 539  &   2,259               & 0.92$\pm$0.22 & 2.13$\pm$0.30 & 0.550$\pm$0.002 & 4.390$\pm$0.005 & 1.44$\pm$0.27 & 2.48$\pm$0.26 & 0.494$\pm$0.003 & 4.383$\pm$ 0.004 \\
    HiFiTTS, LibriTTS & 822 & 2,265 & 0.77$\pm$0.13 & 1.72$\pm$0.17 & 0.723$\pm$0.002 & 4.461$\pm$0.013 & 0.80$\pm$0.07 & 1.67$\pm$0.11 & 0.588$\pm$0.002 & 4.414$\pm$0.011\\
    % HiFiTTS, LibriTTS(1e-5) & 822 & 2,265 & 0.61$\pm$0.07 & 1.64$\pm$0.13 & 0.607$\pm$0.002 & 4.389$\pm$0.006 & 1.29$\pm$0.26 & 2.25$\pm$0.27 & 0.502$\pm$0.002 & 4.398$\pm$0.009\\
    \midrule
    HiFiTTS-2 & 30,400     &   4,940 & 0.47$\pm$0.11 & 1.21$\pm$0.12 & 0.714$\pm$0.002 & 4.388$\pm$0.004 & 0.57$\pm$0.11 & 1.42$\pm$0.13 & 0.731$\pm$0.001 & 4.387$\pm$0.005\\
    HiFiTTS, LibriTTS, HiFiTTS-2 & 31,222 & 5,657 & 0.62$\pm$0.15 & 1.43$\pm$0.17 & 0.739$\pm$0.002 & 4.385$\pm$0.007 & 0.57$\pm$0.04 & 1.39$\pm$0.08 & 0.739$\pm$0.002 & 4.382$\pm$0.006\\
    \bottomrule
    \end{tabular}
    }
    \vspace{-4mm}
    \label{tab:tts_results_final}
\end{table*}

\section{Experiments}

%To validate the advantages of using HiFiTTS-2 for TTS over existing LibriTTS and HiFiTTS from LibriVox family, we picked up one of the state-of-the-art encoder-decoder architectures, decoder-context Koel-TTS~\cite{hussain2025koelttsenhancingllmbased}, for comparative study. The trainable model parameter size is around \texttt{378M}. 

To demonstrate the benefit of using HiFiTTS-2 for speech synthesis, we employ Koel-TTS~\cite{hussain2025koelttsenhancingllmbased}, a state-of-the-art encoder-decoder architecture. We compare performance at 22.05 kHz against existing LibriVox datasets: LibriTTS and HiFiTTS.

\subsection{Koel-TTS Model}
\begin{figure}[htbp]
    % \vspace{-2mm}
    \centering
    \includegraphics[width=0.65\linewidth]{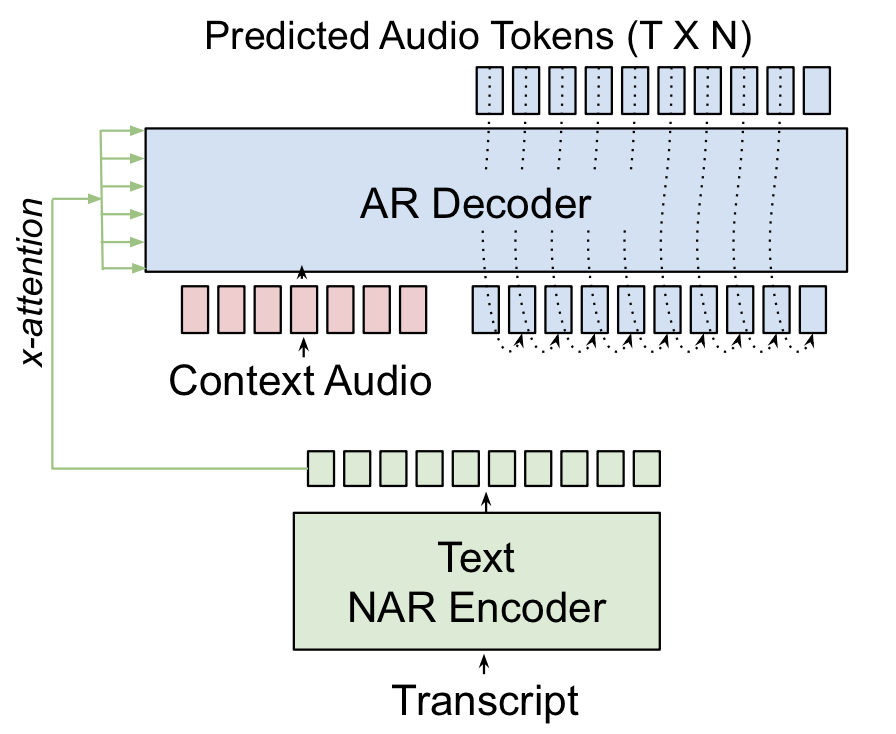}
    \caption{Koel-TTS Model Architecture}
    \vspace{-4mm}
    \label{fig:koeltts}
\end{figure}

Koel-TTS~\cite{hussain2025koelttsenhancingllmbased} consists of an autoregressive~(AR) transformer decoder conditioned on text encodings 
from a non-autoregressive~(NAR) transformer encoder, using cross-attention layers. The model has approximately 378M trainable parameters.
The AR transformer predicts audio tokens frame by frame, generating all codebooks in parallel at each time step, conditioned on the input text and previous audio tokens. As shown in Figure~\ref{fig:koeltts}, the context audio tokens are directly provided as input to the AR decoder by prepending them to the target audio tokens. A single unified transformer decoder processes both the context and target audio tokens, leveraging a shared representation for conditioning and prediction. We adapted classifier-free guidance~(CFG)~\cite{hussain2025koelttsenhancingllmbased} to train models by dropping out the text and audio conditionals with a probability of 10\%. During inference, we applied a CFG scale of 2.5 to guide the AR token prediction for more precise control over the generated speech. 

Our models are trained on 32 NVIDIA DGX H100 GPUs using a global batch size of 512. We use an Adam optimizer with an initial learning rate \texttt{1e-5} for small-scale training with LibriTTS and HiFiTTS, and \texttt{2e-4} for large-scale training with HiFiTTS-2 and their combinations. No weight decays are applied. The learning rate is annealed every 1,000 training steps with an exponential decay factor of 0.998.

\subsection{Evaluation Metrics and Data}
We evaluated the synthesized speech on intelligibility, speaker similarity, and audio quality. Intelligibility is measured with character error rate~(CER) and word error rate~(WER) computed by a state-of-the-art ASR Model Parakeet-TDT~\cite{xu2023efficient}.
%\emph{Token-and-Duration Transducer} ASR model~\cite{xu2023efficient} 
%\footnote{https://huggingface.co/nvidia/parakeet-tdt-1.1b}
Speaker similarity is measured with cosine similarity between synthesized and context audio embeddings extracted by a speaker recognition model Titanet-Small~\cite{koluguri2022titanet}.
%\footnote{https://catalog.ngc.nvidia.com/orgs/nvidia/teams/nemo/models/titanet\_small}
%Naturalness is measured by mean opinion score estimated by \emph{SQUIM-MOS}~\cite{kumar2023torchaudio}.
Audio quality is measured with a reference-free MOS estimator SQUIM-MOS~\cite{kumar2023torchaudio}.
%\footnote{https://pytorch.org/audio/main/tutorials/squim\_tutorial.html}
We use multinomial top-k
sampling with \texttt{k=80} and \texttt{temperature=0.6} during inference, and report mean metrics with 95\% confidence intervals after running 10 times for each experiment.

For \textbf{training data}, we train one model for each individual dataset: LibriTTS, HiFiTTS,
%\footnote{We include both \emph{clean} and \emph{other} subsets from LibriTTS and HiFiTTS.}, 
and HiFiTTS-2. Both \emph{clean} and \emph{other} subsets are chosen for LibriTTS and HiFiTTS. We also experiment with combining datasets. For each dataset, we manually create triplets of (5-second context audio, transcript, target audio), where context and target audio are distinct utterances from the same speaker. For HiFiTTS-2, we also discarded the triplets where target audio CER is greater than 3\% or speaker similarity between context and target audio is less than 0.6. This filtering results in 30,400 hours of training data.

For \textbf{test data}: We evaluate models on subsets of LibriTTS for both seen and unseen speakers. We withhold 200 utterances from 170 speakers in \emph{train-clean-360} as seen speakers, and 180 utterances from 36 speakers in \emph{test-clean} as unseen speakers. We use 5 distinct context and target audio pairs from each speaker.

\subsection{Results and Analysis}
Table~\ref{tab:tts_results_final} illustrates the comparison between Koel-TTS models trained on individual datasets and their combination.
We observed that the model trained on LibriTTS serves as a strong baseline, performing well on metrics for both seen and unseen speakers. As expected, the model trained solely on HiFiTTS underperforms on unseen speakers due to its limited speaker diversity covering only 10 speakers, as shown in Table~\ref{table:stats}. 
%It is expected that a model trained on only 10 speakers will not generalize well to unseen speakers.
Combining LibriTTS and HiFiTTS improves performance on both seen and unseen speakers, achieving performance on SQUIM-MOS slightly better than even the large-scale trainings. However, the most substantial improvements come from training with HiFiTTS-2, which significantly enhances performance, especially in speaker similarity for unseen speakers. This highlights the benefits of our proposed large-scale dataset. Furthermore, training with all three available datasets led to additional improvements, particularly for unseen speakers, emphasizing the critical role of diverse, large-scale data for zero-shot TTS.

%While HiFiTTS-2 excels in most metrics, there's a slight decrease in SQUIM-MOS compared to smaller datasets, indicating potential challenges in maintaining audio quality with increased diversity. Overall, these results demonstrate the effectiveness of the HiFiTTS-2 dataset and our data processing pipeline, particularly in enhancing generalization to unseen speakers while maintaining high audio quality.

\section{Conclusion}

In this paper we introduce HiFiTTS-2, a large-scale English dataset designed for modeling high-bandwidth speech. Derived from LibriVox audiobooks originally downloaded at 48 kHz, the dataset is processed through our custom pipeline to assess and ensure the quality of both audio and transcripts. Our experiments demonstrate the effectiveness of HiFiTTS-2 and our data processing methods for TTS applications, and we believe its utility can extend to other domains such as bandwidth extension. In future, we aim to expand our work to incorporate data from diverse sources and languages, and to explore applications beyond TTS that require high-bandwidth audio processing.

% \section{Acknowledgment}
% \xuesong{TBD, don't add this section when being under review. This can be placed at 5-th page along with reference. So far as I know, Sasha Meister and Evalina were attempting the similar effort back to March 2023 and shared their repo https://gitlab-master.nvidia.com/ebakhturina/tts\_eng\_dataset; we also started this work following the very early initiatives driven by Rafael and Boris https://docs.google.com/document/d/13fWQzB5XA91EqWCoiTok-4jx-QLyRX0QtXeqohT4Ifk/edit?tab=t.0}

\vfill\pagebreak

\bibliographystyle{IEEEtran}
\bibliography{mybib}

\end{document}